\documentclass[aps,pra,showpacs,preprint,groupedaddress]{revtex4}
\usepackage{amssymb,bm}
\usepackage[dvips]{graphicx}

\begin{document}
\bibliographystyle{apsrev}

\title{Polarization effects in non-relativistic $ep$ scattering}

\author{A.I. Milstein  }
\author{S.G. Salnikov  }
\author{V.M.Strakhovenko}
 \affiliation{Budker Institute of Nuclear Physics, 630090 Novosibirsk, Russia}

\date{\today}

\begin{abstract}
The cross section which addresses the spin-flip transitions of a
proton (antiproton) interacting with a polarized non-relativistic
electron or positron is calculated analytically. In the case of
attraction, this cross section is greatly enhanced for sufficiently
small relative velocities as compared to the result obtained in the
Born approximation. However, it is still very small, so that the
beam polarization time turns out to be enormously large for the
parameters of $e^{\pm}$ beams available now. This practically rules
out a use of such beams to polarize stored antiprotons or protons.
\end{abstract}

\maketitle
\section{Introduction}
An extensive physical  program with polarized antiprotons has been
proposed recently by the PAX Collaboration \cite{PAX05}. This
program  has initiated a  discussion of various  methods to polarize
stored antiprotons (see \cite{PAX05, Rathman05} and literature
therein). One option, which was considered, is to use for this
purpose the interaction with a polarized electron (positron) beam.
When the relative velocity  $v$ of two beams is sufficiently small,
the scattered antiprotons remain in the beam. In this case, the
polarization buildup is completely due to the spin-flip transitions,
as it was explained in Refs. \cite{MilStr05,Kolya06}. The
corresponding cross section was estimated in Ref. \cite{MilStr05} in
the Born approximation and turned out to be too small (about
$10^{-3}\, \mbox{\rm mb}$) to provide a noticeable polarization in a
reasonable time.

At very small velocities, when $\alpha/v\gtrsim 1$ ($\alpha$ is the
fine-structure constant and $\hbar=c=1$), the Born approximation
becomes invalid. It is well known that for such velocities the cross
section is modified due to a corresponding change of the $e^+$
($e^-$) flux near the origin. A magnitude of the effect can be
characterized (see, e.g. \cite{LL}) by the parameter $C(\xi)$ being
the squared ratio of the Coulomb and free wave function at the
origin:
\begin{equation}\label{CC}
C(\xi)= \frac{2\pi\xi}{\exp(2\pi\xi)-1}\,,
\end{equation}
here $\xi = e_1 e_2/v$, so that $\xi = \alpha/v$ for repulsion ($e^+
p$ or $e^- \bar p$ interaction) and $\xi = -\alpha/v$ for attraction
($e^- p$ or $e^+ \bar p$ interaction). At $|\xi|\gg 1$, the quantity
$C(\xi)$ is large only in the case of attraction where $C(\xi) =
2\pi|\xi|$. Since the cross section contains terms $\propto
C^2(\xi)$, we can figure on the enhancement in $e^- p$ or $e^+ \bar
p$ interaction of the order $(2\pi\xi)^2$ as compared with the Born
approximation.

Recently, the enhancement that exceeds $C^2(\xi)$ by many orders of
magnitude was claimed in Refs. \cite{WA,Aren}. Correspondingly, the
conclusion was drawn in \cite{WA} that antiprotons can be easily
polarized by interaction with a polarized positron beam. The crucial
point of the calculation, estimation of the radial integrals, was
performed in \cite{WA,Aren} numerically.

To shed light on the problem, we perform in the present paper an
analytical calculation of the cross section, which addresses the
spin-flip transitions of an antiproton (proton) interacting with a
polarized non-relativistic positron or electron. Just $C^2(\xi)$ is
obtained as the enhancement factor. It is big enough, but the
resulting cross section is still very small (less than $1\,
\mbox{\rm mb}$ at $v$=0.0019 used for estimations in \cite{WA}).
Using this cross section, we analyze the kinetics of the
polarization buildup. The polarization time turns out to be
enormously large for the parameters of $e^+$ or $e^-$ beams
available now. To give a feeling of how far we are from the
practical use of the effect, we note that the scale of days for the
polarization time at $v$=0.0019 can be achieved when the positron
beam density becomes higher than $5\cdot 10^{16} \mbox{\rm
cm}^{-3}$.

\section{Cross section}
We assume that  $v\ll 1$, where $v$ is the relative velocity of an
electron and a proton. Then one can neglect recoil effects for the
proton and consider it as a source of the Coulomb field. Thus, we
calculate all spin effects with the help of perturbation theory
using the exact Coulomb non-relativistic wave functions. A part of
the Hamiltonian of the $ep$ interaction dependent on the proton spin
reads \cite{BLP}
\begin{eqnarray}\label{H}
H=\frac{\alpha\mu_0}{m_em_p}\Bigg\{\frac{3(\bm n\cdot\bm s_1)(\bm
n\cdot\bm s_2)-(\bm s_1\cdot\bm s_2)}{r^3} +\frac{8\pi}{3}\delta(\bm
r)(\bm s_1\cdot\bm s_2)+\frac{\bm L\cdot\bm s_1}{r^3}\Bigg\}\,,
\end{eqnarray}
where  $\bm s_1$ and $\bm s_2$  are the spin operators of a proton
and an electron, respectively, $\bm L$ is the angular momentum
operator of an electron, and $\mu_0=2.79$ is the dimensionless
proton magnetic moment in units of the nuclear magneton. We
introduce the spinors $\phi_\sigma$ and  $\chi_\lambda$ describing
the proton and electron spin states, respectively. Let the initial
electron be polarized along the unit vector $\bm \zeta_e$ and the
initial proton along $\bm \zeta_p$, so that $2\phi_{\sigma}^+\bm
s_1\phi_\sigma=\bm \zeta_p$ and $2\chi_{\lambda_i}^+\bm s_2
\chi_{\lambda_i}=\bm \zeta_e$. Then the cross section of scattering
with the spin-flip transition of the proton has the form
\begin{eqnarray}\label{dsigma}
d\sigma= \frac{m_e^2d\Omega_{2}}{(2\pi)^2}\sum_{\lambda_f} |M|^2\,,
\quad M=\int d\bm r\,\phi_{-\sigma}^+ \chi_{\lambda_f}^+
\psi^{*(-)}_{\bm p_2}(\bm r) H\psi^{(+)}_{\bm p_1}(\bm r)\phi_\sigma
\chi_{\lambda_i}
\end{eqnarray}
where $\bm p_1$ and  $\bm p_2$ are the momenta of the initial and
final electron, $\Omega_2$ is the solid angle of the vector $\bm
p_2$, $\psi^{(+)}_{\bm p_1}(\bm r)$ is the wave function of the
initial electron containing at large distances a plain wave and the
divergent spherical wave, $\psi^{(-)}_{\bm p_2}(\bm r)$ is the wave
function of the final electron containing at large distances  a
plain wave and the convergent spherical wave. These wave functions
read
\begin{eqnarray}\label{psi}
&&\psi^{(-)}_{\bm p_2}(\bm r)=\frac{1}{2p}\sum_{l=0}^{\infty}(2l+1)i^l
\mbox{e}^{-i\delta_l}P_l(\bm k_2\cdot\bm n)R_{p,l}(r)
\,,\nonumber\\
&&\psi^{(+)}_{\bm p_1}(\bm r)=\frac{1}{2p}\sum_{l=0}^{\infty}(2l+1)i^l
\mbox{e}^{i\delta_l}P_l(\bm k_1\cdot\bm n)R_{p,l}(r)
\,.
\end{eqnarray}
Here $\bm k_{1,2}=\bm p_{1,2}/p$ , $\bm n=\bm r/r$ , $\delta_l$ and
$R_{p,l}(r)$ are, respectively, the phase shifts and radial wave
functions of an electron in a Coulomb field, $P_l(x)$  are the
Legendre polynomials. Substituting these wave functions into
Eq.~(\ref{dsigma}) we obtain the matrix element in the form
\begin{eqnarray}\label{M}
&&M=\frac{\alpha\mu_0}{4p^2m_em_p}\Bigg\{\frac{8\pi}{3}(\bm
S_1\cdot\bm S_2)\mbox{e}^{2i\delta_0}R_{p,0}^2(0)+
\sum_{l=1}^{\infty}(2l+1)^2\mbox{e}^{2i\delta_l}\int\limits_0^{\infty}
\frac{dr}{r}R_{p,l}^2(r)\nonumber\\
&&\times\int d\bm n P_l(\bm k_2\cdot\bm n)[3(\bm n\cdot\bm S_1)(\bm
n\cdot\bm S_2)-(\bm S_1\cdot\bm S_2) + \delta_{\lambda_f,\lambda_i}
\bm L\cdot\bm S_1]
P_l(\bm k_1\cdot\bm n)\nonumber\\
&&-\sum_{l=0}^{\infty}(2l+1)(2l+5)\mbox{e}^{i(\delta_l+\delta_{l+2})}
\int\limits_0^{\infty}
\frac{dr}{r}R_{p,l}(r)R_{p,l+2}(r)\nonumber\\
&&\times\int d\bm n[P_l(\bm k_2\cdot\bm n)P_{l+2}(\bm k_1\cdot\bm n)+
P_l(\bm k_1\cdot\bm n)P_{l+2}(\bm k_2\cdot\bm n)]\nonumber\\
&&\times[3(\bm n\cdot\bm S_1)(\bm n\cdot\bm S_2)-(\bm S_1\cdot\bm S_2)]\Bigg\}\, ,
\end{eqnarray}
where $\bm S_1=\phi_{-\sigma}^+ \bm s_1\phi_{\sigma}$ and $\bm
S_2=\chi_{\lambda_f}^+\bm s_2\chi_{\lambda_i}$.

In what follows we assume that in the rest frame of a proton beam
the velocities $\bm v$ of electrons are randomly distributed over
their directions. Therefore we average the cross section over the
direction of the vector $\bm k_1$. It is convenient to perform
integrations in the following order. We integrate $|M|^2$ first over
$\bm k_1$ and  $\bm k_2$ and then over the angles of $\bm n$ using
the relations
\begin{eqnarray}\label{rel}
&&\int d\bm k P_l(\bm k\cdot\bm n) P_l(\bm k\cdot\bm n')=\frac{4\pi}{2l+1}P_l(\bm n\cdot\bm n')\, ,\nonumber\\
&&\int\!\!\!\int  d\bm n_1  d\bm n_2 P_l(\bm n_1\cdot \bm n_2) P_{l'}(\bm n_1\cdot \bm n_2)(3n_1^in_1^j-\delta^{ij})(3n_2^an_2^b-\delta^{ab})\nonumber\\
&&=\frac{8\pi^2}{5}[3(\delta^{ia}\delta^{jb}+\delta^{ib}\delta^{ja})-2\delta^{ij}\delta^{ab}]\int\limits_{-1}^1
dx P_2(x)P_l(x)P_{l'}(x)\, ,\nonumber\\
&&\int\!\!\!\int  d\bm n_1  d\bm n_2 [(\bm S_1\cdot \bm L_1) P_l(\bm n_1\cdot \bm n_2)]
[(\bm S_1^*\cdot \bm L_2) P_l(\bm n_1\cdot \bm n_2)]=\frac{(4\pi)^2}{3}(\bm S_1\cdot \bm S_1^*)\frac{l(l+1)}{2l+1}\, .
\end{eqnarray}
As a result we have
\begin{eqnarray}\label{sigma}
&&\sigma=\pi\left(\frac{\alpha\mu_0}{p^2m_p}\right)^2\sum_{\lambda_f
}\Bigg\{ \frac{1}{5}\Big[3\Big((\bm S_1\cdot \bm S_1^*) (\bm
S_2\cdot \bm S_2^*)+
|(\bm S_1\cdot \bm S_2^*)|^2\Big)-2|(\bm S_1\cdot \bm S_2)|^2\Big]\nonumber\\
&&\times\Big[\sum_{l=1}^\infty \frac{l(l+1)(2l+1)}{(2l+3)(2l-1)}{\cal F}_{l}^2+
\sum_{l=0}^\infty \frac{3(l+1)(l+2)}{(2l+3)}{\cal G}_{l}^2\Big]+
\left|\frac{2}{3}(\bm S_1\cdot \bm S_2)R^2_{p,0}(0)\right|^2\nonumber\\
&&+\frac{1}{3} \delta_{\lambda_f,\lambda_i}(\bm S_1\cdot \bm
S_1^*)\sum_{l=1}^\infty l(l+1)(2l+1){\cal F}_{l}^2\Bigg\}\,
\end{eqnarray}
with
\begin{eqnarray}\label{fg}
{\cal F}_{l}= \int\limits_0^{\infty}\frac{dr}{r}R_{p,l}^2(r)\,
,\quad {\cal G}_{l}= \int\limits_0^{\infty}\frac{dr}{r}R_{p,l}(r)
R_{p,l+2}(r)\, .
\end{eqnarray}
Calculating the vectors $\bm S_1$ and $\bm S_2$ and taking the sum
over $\lambda_f$, we finally arrive at
\begin{eqnarray}\label{sigmaf}
&&\sigma=\pi\left(\frac{\alpha\mu_0}{p^2m_p}\right)^2\Bigg\{
\frac{1}{8}(2+\bm \zeta_e\cdot \bm \zeta_p)\Big[\sum_{l=1}^\infty \frac{l(l+1)(2l+1)}{(2l+3)(2l-1)}{\cal F}_{l}^2
+\sum_{l=0}^\infty \frac{3(l+1)(l+2)}{(2l+3)}{\cal G}_{l}^2\Big]  \nonumber\\
&&+\frac{1}{18}(1-\bm \zeta_e\cdot \bm \zeta_p)R^4_{p,0}(0)
+\frac{1}{6}\sum_{l=1}^\infty l(l+1)(2l+1){\cal F}_{l}^2\Bigg\}\,.
\end{eqnarray}
A contribution  of the $LS$-interaction to $\sigma$ is given by the
last term in Eq. (\ref{sigmaf}) which is independent of the electron
spin. Therefore, it does not lead to the appearance of a hadron beam
polarization but contributes to its depolarization.
\section{Radial integrals}
Let us consider the  functions ${\cal F}_{l}$ and ${\cal G}_{l}$. We
calculate them using the convenient integral representation of the
electron Green function in a Coulomb field found in Ref.
\cite{MSo21} and the relation for the wave functions of the
continuous spectrum
\begin{equation}\label{deltaG}
\int \frac{d\bm k}{4\pi}\, \psi_{\bm p}(\bm r_1)\psi^{*}_{\bm p}(\bm r_2)=
\frac{i\pi}{pm_e}\delta G(\bm r_1,\,\bm r_2|E)\, ,
\end{equation}
where $\bm k=\bm p/p$ , $E=p^2/(2m_e)$, and
$$\delta G(\bm r_1,\,\bm r_2|E)=G(\bm r_1,\,\bm r_2|E+i0)  - G(\bm r_1,\,\bm r_2|E-i0)$$
is the discontinuity of the Green function on the cut. Then we obtain
\begin{eqnarray}\label{RRG}
 R_{p,l}(r_1)R_{p,l}(r_2)=\frac{2p(-1)^{l+1}}{\sqrt{r_1r_2}}\int\limits_{-\infty}^{+\infty}\frac{ds}{\sinh s}
\exp\left[ip(r_1+r_2)\coth s -2i\xi
s\right]\,J_{2l+1}\left(\frac{2p\sqrt{r_1r_2}}{\sinh s}\right)\, ,
\end{eqnarray}
where $J_\nu (x)$ is the Bessel function. Remember that, according
to our definition Eq. (\ref{CC}), $\xi=-\alpha/v$ for the attractive
potential and $\xi=\alpha/v$ for the repulsive one. For the square
of the radial wave function we obtain from Eq. (\ref{RRG})
\begin{eqnarray}\label{R2}
 R_{p,l}^2(r)=\frac{2p(-1)^{l+1}}{r}\int\limits_{-\infty}^{+\infty}\frac{ds}{\sinh s}
\exp\left[2ipr\coth s -2i\xi
s\right]\,J_{2l+1}\left(\frac{2pr}{\sinh s}\right)\, .
\end{eqnarray}
Substituting the asymptotic form of $R_{p,l}(r_2)$ at small
distances \cite{LL} into Eq. (\ref{RRG}), we arrive at the following
integral representation for $R_{p,l}(r)$
\begin{eqnarray}\label{R1}
 R_{p,l}(r)=\frac{p\,2^{-l}(pr)^l(-1)^{l+1}}{\exp(-\pi\xi/2)|\Gamma(l+1+i\xi)|}
\int\limits_{-\infty}^{+\infty}\frac{ds}{(\sinh s)^{2l+2}}
\exp\left[ipr\coth s -2i\xi s\right],
\end{eqnarray}
where $\Gamma(x)$ is the Euler gamma function. Then we substitute
Eq. (\ref{R2}) into the definition of the function $ {\cal F}_{l}$
in Eq. (\ref{fg})  and take the integral first over $r$ and then
over the parameter $s$. In order to change the order of integration
we deform the contour of integration over $s$ so that it passes in
the positive direction  around the point $s=0$. After that all
integrals can be easily taken and we obtain
\begin{eqnarray}\label{calf}
{\cal F}_{l} =\frac{(2p)^2}{2l(l+1)}f_l(\xi)\,,\quad f_l(\xi)=
\left\{1+\frac{\xi}{2l+1}
\left[2\mbox{Im}\,\psi(l+1+i\xi)-\pi\right]\right\}\,,
\end{eqnarray}
where $\psi(z)=d\ln\Gamma(z)/dz$. In the limit $\xi\rightarrow 0$,
i.e. in the Born approximation, $f_l(\xi)\rightarrow 1$. The term in
$f_l(\xi)$ linear with respect to $\xi$ is the only odd term. It
comes from the contribution of the small half-circle around the
point $s=0$. All the higher order terms in $\xi$ are the even
functions of $\xi$. Using the relation
$$\mbox{Im}\,\psi(l+1+i\xi)=\xi \sum_{k=0}^\infty
\frac{1}{(k+l+1)^2+\xi^2}\,,$$ we can present the function
$f_l(\xi)$ as
\begin{equation}\label{ffl}
f_l(\xi)= \frac{1}{2l+1}\left[C(\xi)+2\sum_{k=1}^l
\frac{k^2}{k^2+\xi^2}\right]\,,
\end{equation}
where $C(\xi)$ is just the factor introduced in Eq. (\ref{CC}). So,
at $|\xi|\gg 1$ the radial integral ${\cal F}_{l}$ is enhanced by
the factor $2\pi|\xi|$ for the attractive potential ($\xi<0$). For
the repulsive potential ($\xi>0$), it is suppressed as $1/\xi^2$,
not exponentially as one can na\"{\i}vely  expect. The
straightforward calculation with the use of Eq. (\ref{R1}) leads to
the following expression for the function ${\cal G}_{l}$ Eq.
(\ref{fg})
\begin{eqnarray}\label{calg}
{\cal G}_{l} =\frac{(2p)^2}{6|l+1+i\xi||l+2+i\xi|}\,.
\end{eqnarray}
Note that ${\cal G}_{l}$ is an even function of $\xi$, being
suppressed at $|\xi|\gg l$ as $1/\xi^2$.

Substituting the calculated radial integrals Eqs. (\ref{calf}) and
(\ref{calg}) into Eq. (\ref{sigmaf}) and recollecting that
$R^4_{p,0}(0)= (2p)^4 C^2(\xi)$, we obtain the final form of the
cross section
\begin{eqnarray}\label{crfin}
&&\sigma=\frac{1}{2}\sigma_0\Bigg\{C^2(\xi) \Big[(3-2\ln 2)+(\ln
2-2)(\bm \zeta_e\cdot \bm \zeta_p)\Big]+\sum_{l=1}^\infty
\frac{1}{l(l+1)(2l+1)}\Bigg[4 C(\xi)g_l \nonumber\\
&&+(2+\bm \zeta_e\cdot \bm
\zeta_p)\Bigg(\frac{3(C(\xi)+g_l)g_l}{(2l-1)(2l+3)}
+\frac{l^2(l+1)^2}{4(l^2+\xi^2)((l+1)^2+\xi^2)}\Bigg)\Bigg]\\&&+\sum_{l=1}^{l_{max}}
\frac{(2g_l)^2}{l(l+1)(2l+1)}\Bigg\}\,. \nonumber
\end{eqnarray}
Here $$\sigma_0=\frac{\pi}{3}\left(\frac{2\alpha\mu_0}{m_p}\right)^2
\thickapprox 0.77\cdot10^{-3}\mbox{mb}\,, \quad g_l=\sum_{k=1}^l
\frac{k^2}{k^2+\xi^2}\,.$$ The last sum in Eq. (\ref{crfin}) has
been regularized by  introducing $l_{max}\gg 1$, as it is
logarithmically divergent at large $l$. This divergence is well
known (see, e.g. \cite{BLP}) and addresses a singular behavior of
the Born amplitude at small scattering angle. So that we can set
$l_{max}=p\rho$, where $\rho$ has a meaning of a maximal impact
parameter being of the order of the transverse beam size. At $v\ll
1$, the expression Eq. (\ref{crfin}) is valid for any values of
$\xi$. At $\xi\rightarrow 0$ it goes over into
$$\sigma_{Born}=\sigma_0[1-\frac{1}{2}(\bm \zeta_e\cdot \bm
\zeta_p)+\ln(l_{max})]\,,$$ where the spin independent term
dominates. It is evident from Eq. (\ref{crfin}) that we can figure
on some polarization effects only for the attractive potential
($\xi<0$) at $|\xi|\gg 1$. For this case, we obtain from Eq.
(\ref{crfin})

\begin{eqnarray}\label{sigmaas}
\sigma=\sigma_0\Bigg\{ \frac{1}{2}(2\pi\xi)^2 \Bigg[(3-2\ln 2)+(\ln
2-2)(\bm \zeta_e\cdot \bm \zeta_p)\Bigg]+\ln\frac{
l_{max}}{|\xi|}\Bigg\}\,.
\end{eqnarray}
Note that typically $l_{max}\gg |\xi|$ which has  been used in the
derivation of Eq. (\ref{sigmaas}).

\section{Kinetics of polarization}
The kinetics of the polarization buildup due to interaction with
various polarized targets was considered in detail in Ref.
\cite{MilStr05}. In the case of initially unpolarized proton beam
interacting with the polarized electron beam, when the direction of
the relative velocity $\bm v$ of electron and proton is random, the
arising polarization of the proton beam is collinear with the
direction of the electron beam polarization. The polarization degree
of the proton beam , $P(t)$ , at time $t$ reads
\begin{eqnarray}\label{solution}
&&P(t)=P_e P_0\left(1-\mbox{e}^{-\Omega t}\right)\,,\quad P_0=
\frac{\sigma_{+-}-\sigma_{-+}}{\sigma_{+-}+\sigma_{-+}}\, ,\nonumber\\
&& \Omega=fnl\sigma_{tot}\frac{v}{V_b}\, ,\quad \quad \quad
\sigma_{tot}=\sigma_{+-}+\sigma_{-+}\, .
\end{eqnarray}
Here $f$ is a revolution frequency, $n$ is a density of the electron
beam, $l$ is the length of the interaction region, $V_b$ is the
proton beam velocity. All these quantities are defined in the lab
frame. In Eq. (\ref{solution}), $P_e$ is the polarization degree of
the electron beam  and the cross sections $\sigma_{-+}$ and
$\sigma_{+-}$ are obtained from Eq. (\ref{sigmaas}) with $\bm
\zeta_p=\bm \zeta_e$  and $\bm \zeta_p=-\bm \zeta_e$, respectively.
If $P_e=0$ but the proton beam has some initial polarization $P(0)$,
then $P(t)=P(0)\exp(-\Omega t)$, i.e. the rates of depolarization
and polarization are the same.

From Eq. (\ref{sigmaas}) we obtain
\begin{eqnarray}\label{pomega}
P_0 &=&\frac{(2-\ln 2)}{
3-2\ln 2+\ln (l_{max}^2/\xi^2)/(2\pi\xi)^2}\,,\nonumber\\
\sigma_{tot} &=&\sigma_0\Bigg\{ (2\pi\xi)^2 (3-2\ln 2)+\ln\frac{
l_{max}^2}{\xi^2}\Bigg\}\,.
\end{eqnarray}
At $v$=0.0019 used for estimations in \cite{WA} and $\rho\approx
1$~cm , we have  $P_0\approx 0.78$ which is big enough. However, for
the same parameters, we have $\sigma_{tot}\approx 0.75\,\mbox{mb}$
that is drastically different from the result $\sigma_{tot}\approx
4\cdot 10^{+13}\,\mbox{barn}$ obtained in \cite{WA}. When the
correct value of $\sigma_{tot}$ is used, the beam polarization time,
$\tau=\Omega^{-1}$, becomes enormously large for the parameters of
$e^{\pm}$ beams available now. For example, to reach the scale of
days for $\tau$ at values of $v$, $f$, and $l$ used in \cite{WA},
the positron beam density should be higher than $n\approx
5\cdot10^{16} \mbox{\rm cm}^{-3}$. Even in the case of the permanent
interaction of two beams, when $\tau=1/(nv\sigma_{tot})$ is minimal,
the polarization time about one day is obtained at $v$=0.0019 for
the density $n\approx 3\cdot10^{14} \mbox{\rm cm}^{-3}$.

In the present paper, the cross section which is averaged over
directions of relative velocities (over $\bm k_1=\bm p_1/p$), is
calculated. We have also obtained the cross section ${\tilde
\sigma}$ without such averaging.  At $\xi<0$ and $|\xi|\gg 1$,  we
have for ${\tilde \sigma}$

\begin{eqnarray}\label{sigmanon}
{\tilde \sigma}=\sigma&+&\frac{\sigma_0}{8}\Biggl\{ [3(\bm
\zeta_p\cdot \bm k_1 )^2-1][ (2\pi\xi)^2(\ln
2-\frac{1}{2})+\ln\frac{ l_{max}^2}{\xi^2}]\nonumber\\\\
&+&(2\pi\xi)^2(2\ln 2-1)[(\bm \zeta_e\cdot \bm \zeta_p)-3(\bm
\zeta_e\cdot \bm k_1)(\bm \zeta_p\cdot \bm k_1 )]\Biggr\}
\,,\nonumber
\end{eqnarray}
where $\sigma$ is the cross section defined by Eq. (\ref{sigmaas}).
From Eq. (\ref{sigmanon}), we obtain for the cross section ${\tilde
\sigma}_{tot}={\tilde \sigma}_{+-}+{\tilde \sigma}_{-+}$ in the case
of the transverse ($\bm \zeta_e \perp \bm k_1$) and longitudinal
($\bm \zeta_e
\parallel \bm k_1$) polarization
\begin{eqnarray}\label{sigatot}
{\tilde \sigma}_{tot}^{\perp}&=&\frac{3}{4}\sigma_0\Bigg\{
(2\pi\xi)^2(\frac{25}{6}-3\ln 2)+\ln\frac{ l_{max}^2}{\xi^2}
\Bigg\}\,,\nonumber\\\\
{\tilde \sigma}_{tot}^{\parallel}&=&\frac{3}{2}\sigma_0\Bigg\{
(2\pi\xi)^2 (\frac{11}{6}-\ln 2)+\ln\frac{
l_{max}^2}{\xi^2}\Bigg\}\,.\nonumber
\end{eqnarray}
In particular, at $v$=0.0019 and $\rho\approx 1$~cm , we have
${\tilde \sigma}_{tot}^{\perp}\approx 0.72\,\mbox{mb}$ and ${\tilde
\sigma}_{tot}^{\parallel}\approx 0.81\,\mbox{mb}$.  These values are
not very different from each other and from $\sigma_{tot}\approx
0.75\,\mbox{mb}$ since a scale of the cross section is determined by
values of radial integrals in the matrix element (\ref{M}). Let us
note that the contribution of the logarithmic term to the cross
section at $|\xi|\gg 1$ is relatively small being about several
percent in the above examples. If we neglect such terms, the
dependence of the cross section  on relative velocity is reduced to
the common factor $\xi^2$.

We conclude that the cross section, which addresses the spin-flip
transitions of an antiproton (proton) interacting with a polarized
non-relativistic positron or electron is derived analytically. In
the case of attraction, the cross section is strongly enhanced at
$\alpha/v \gg 1$ as compared with that obtained in the Born
approximation. However, this enhancement is insufficient for
providing a practical use of positron beams to polarize stored
antiprotons.  Thus, the filtering method is still the most promising
way for that (see, e.g. \cite {MSS08}).


\begin{thebibliography}{99}

\bibitem{PAX05} Technical Proposal for Antiproton-Proton Scattering Experiments with
Polarization, PAX Collaboration,  arXiv:hep-ex/0505054 (2005).

\bibitem{Rathman05} F. Rathmann, Current status of the PAX project  (9th PAX Meeting), Dubna,
Russia, September 2005; available from the PAX website at
http://www.fz-juelich.de/ikp/pax.

\bibitem{MilStr05} A.I.~Milstein and V.M.~Strakhovenko, Phys. Rev. E {\bf 72} (2005) 066503.

\bibitem{Kolya06} N.N.~Nikolaev and F.F.~Pavlov, hep-ph/0601184.

\bibitem{LL} L.D. Landau, E.M. Lifshits,
{\it Quantum Mechanics, Nonrelativistic Theory}, Pergamon, Oxford
(1965).

\bibitem{WA} Th. Walcher, H. Arenh\"ovel, K. Aulenbacher, R. Barday, A. Jankowiak,
Eur. Phys. J. {\bf A 34} (2007) 447.

\bibitem{Aren} H. Arenh\"ovel, Eur. Phys. J. {\bf A 34} (2007) 303.

\bibitem{BLP} V.B. Berestetski, E.M. Lifshits, and L.P. Pitaevsky,
{\it Quantum Electrodynamics}, Pergamon, Oxford (1982).

\bibitem{MSo21} A.I. Milstein, V.M. Strakhovenko, Phys. Lett. {\bf A 92} (1982) 381.


\bibitem{MSS08} V.F. Dmitriev, A.I. Milstein, and V.M. Strakhovenko,
Nucl. Instr. and Meth. {\bf B 266} (2008) 1122.



\end{thebibliography}
\end{document}